\documentclass[12pt]{article}
\usepackage{graphicx} 
\usepackage[utf8]{inputenc}
\usepackage{lscape}
\usepackage{authblk}
\usepackage{float}
\usepackage{mathtools}
\usepackage{longtable}
\usepackage{amsfonts}
\usepackage{amsmath}
 \usepackage{hyperref}
\usepackage{mathabx}
\usepackage{graphicx}
\usepackage{bm}
\usepackage{xcolor}
\usepackage{appendix}
\usepackage[normalem]{ulem}
\usepackage{relsize}
\usepackage{adjustbox}
\usepackage{setspace}
\usepackage{tabularray}
\usepackage{pdflscape}
\usepackage{subfigure}
\usepackage[margin=1in]{geometry}
\newcommand{\ind}{\perp \!\!\! \perp}
 \newcommand{\Acomp}{ \mathcal{A}^{\mathsf{c}}}
\newcommand{\keywords}[1]{\textbf{Keywords:} #1}

\newcommand{\msc}[1]{\textbf{MSC 2020:} #1}

\begin{document}

\title{Bridging Binarization: Causal Inference with Dichotomized Continuous Exposures}
\author[1]{Kaitlyn Lee}
\author[1]{Alan Hubbard}
\author[1]{Alejandro Schuler}
\affil[1]{Division of Biostatistics, University of California, Berkeley}

\maketitle
\begin{abstract}
    The average treatment effect (ATE) is a common parameter estimated in causal inference literature, but it is only defined for binary exposures. Thus, despite concerns raised by some researchers, many studies seeking to estimate the causal effect of a continuous exposure create a new binary exposure variable by dichotomizing the continuous values into two categories. In this paper, we affirm binarization as a statistically valid method for answering causal questions about continuous exposures by showing the equivalence between the binarized ATE and the difference in the average outcomes of two specific modified treatment policies. These policies impose cut-offs corresponding to the binarized exposure variable and assume preservation of relative self-selection. Relative self-selection is the ratio of the probability density of an individual having an exposure equal to one value of the continuous exposure variable versus another. The policies assume that, for any two values of the exposure variable with non-zero probability density after the cut-off, this ratio will remain unchanged. Through this equivalence, we clarify the assumptions underlying binarization and discuss how to properly interpret the resulting estimator. Additionally, we introduce a new target parameter that can be computed after binarization that considers the observed world as a benchmark. We argue that this parameter addresses more relevant causal questions than the traditional binarized ATE parameter. We present a simulation study to illustrate the implications of these assumptions when analyzing data and to demonstrate how to correctly implement estimators of the parameters discussed. Finally, we present an application of this method to evaluate the effect of a law in the state of California which seeks to limit exposures to oil and gas wells on birth outcomes to further illustrate the underlying assumptions.
\end{abstract}


\msc{62D20}
\section{Introduction}

Causal inference methodology historically has considered settings with binary exposure variables. For example, Neyman focused on estimating the average treatment effect (ATE), which compares the average outcome under treatment to the average outcome under no treatment \cite{rubin_neymans_2015}. The ATE continues to be a popular target parameter in causal literature focused on binary exposures. However, across a variety of disciplines, there is growing interest in quantifying the effects of continuous exposures. For example, recent research has explored the causal effects of income inequality \cite{shimonovich_assessing_2022}, alcohol consumption \cite{zhou_alcohol_2022}, and pesticide use \cite{hu_long-_2015}.

The ATE is not defined for continuous exposures; however, one popular method of quantifying the causal effect of a continuous exposure is to binarize or dichotomize the exposure variable, which enables researchers to target a parameter somewhat akin to the ATE. This method involves creating a new dummy variable that equals 1 when the exposure of interest is within some range of values and 0 otherwise. Then, researchers calculate the ATE using this new dummy variable as the exposure. We term this estimand the ``binarized ATE" (BATE).

An example of this practice can be found in a 2015 paper by Hu et al., where the authors investigated the relationship between pesticide exposure and health effects in Chinese farmers \cite{hu_long-_2015}. Individuals were divided into two groups: those with relatively high pesticide exposure levels (more than 50 pesticide applications in 2009--2011) and those with relatively low pesticide exposure levels (less than 50 pesticide applications in the same three year period). The researchers implemented a regression estimator including baseline covariates. They found evidence suggestive of some causal relationship between ``high'' versus ``low'' levels of pesticide use and abnormal health test results (including blood tests, nerve conduction studies, and neurological exams). 

There are numerous other examples of researchers binarizing continuous exposures in public health contexts. For example, Hu et al. investigated the relationship between being in a ``low-risk'' group versus a ``high-risk'' group and risk of type 2 diabetes in women \cite{hu_diet_2001}. Wu et al. calculated the effect of air pollution and mortality \cite{wu_causal_2019}. In addition, researchers employ the binarized ATE across the social sciences, like in political science and economics. Kurtz and Lauretig investigated the effect of economic liberalization on political protest activity \cite{kurtz_does_2022}. Carneiro, Lokshin, and Umapathi looked at the effect of increased schooling on income in Indonesia \cite{carneiro_average_2017}. 

In this paper, we show that the BATE is the same as the difference in average outcomes under two related modified treatment policies (MTPs). MTPs are an example of a population-level intervention in which one considers a counterfactual world in which changes to underlying distributions of the exposure of interest are considered \cite{hubbard_population_2008}. In particular, MTPs are policies that change the underlying distribution of the exposure variable in a population such that the new distribution depends on covariates as well as an individual's natural value of exposure, as defined by Haneuse and Rotnitzky \cite{haneuse_estimation_2013}. 

Utilizing this framework, we show that the binarized ATE  is equivalent to the difference in expected outcome under two policies that impose cut-offs corresponding to the cut-offs used to generate the binarized exposure variable and assume preservation of \textit{relative self-selection}. Relative self-selection is the ratio of the probability density of an individual having an exposure equal to one value of the continuous exposure variable versus another. The policies assume that, for any two values of the exposure variable with non-zero probability density after the cut-off, this ratio will remain unchanged. Equivalently, we can also think of the resulting distribution of exposure as having the distribution of $A$ conditional on truncating the exposure to respect the imposed cut-off.

This equivalence addresses many of the common critiques put forward by methods researchers regarding binarization in causal inference. There have been textbook chapters written on epidemiological methods heavily dissuading researchers from categorizing/binarizing exposure variables \cite{rothman_modern_2008}. Some researchers contend that binarization violates the Stable Unit Treatment Values Assumption (SUTVA), an assumption that some researchers make when prescribing a causal meaning to their statistical estimate of the ATE. In particular, one component of SUTVA is that there are no multiple versions of treatment  \cite{rubin_randomization_1980}. VanderWeele et al. argue that binarization violates this assumption as encoding a continuous exposure into a binary one implies that there are several values of exposure embedded in one value of the binary exposure variable  \cite{vanderweele_inference_2011, vanderweele_causal_2013}. Rothman et al. make a similar argument, stating that such ``open-ended categories," like cut-offs, are harmful because one category can contain many different exposure values or confounder effects \cite{rothman_modern_2008}. However, the proposed MTPs circumvent these issues by defining the precise mechanism by which the distribution of the exposure changes. Specifically, one can re-write the average outcome under two seemingly deterministic binarized exposure levels in terms of the average outcomes under two policies that change the underlying distribution of exposure in a fully specified way. In addition to this critique, Bennette and Vickers contend that binarization assumes ``homogeneity of risk within categories" \cite{bennette_against_2012}. However, the application of MTPs in this paper does not imply that the effect is the same for everyone; rather, in thinking about the effect of a modified treatment policy at a population level, we are considering a causal estimand that is an average of the exposure effect across the distribution of the exposure of interest in the target population.

This type of dichotomization was discussed by Stitelman et al. \cite{stitelman_impact_nodate}. These interventions were also considered by D{\'\i}az and van der Laan as one example of an intervention which imposes some type of cut-off on the distribution of exposure \cite{diaz_munoz_population_2012}. McCoy et al. introduced the ``average regional effect,'' which considers how the average outcome changes under modified treatment policies which preserve relative self-selection \cite{mccoy2023crossvalidated}. Rose and Shem-Tov consider binarization in the context of noncompliance \cite{rose_recoding_2023}. Additionally, the binarized ATE was considered in the context of non-random missingness by van der Laan et al. \cite{van_der_laan_nonparametric_2023}. The estimand in \cite{van_der_laan_nonparametric_2023} reduces to the estimand we present when assuming missingness at random. In this paper, we aim to provide a thorough and detailed treatment of the assumptions underlying the BATE, as well as examples of when the assumptions may be more or less plausible, to provide practitioners with more clarity regarding whether the BATE is the answer to their particular research question. We also present a novel estimand, the causal attributable effect of binarization (CAB), which extends the additive risk estimand considered in \cite{stitelman_impact_nodate}. We argue that this estimand answers a more relevant causal question than the BATE as it compares the average outcome under one MTP to that in the observed world. In addition, we provide details for regression, inverse probability weighted, and augmented inverse probability weighted estimators that, under certain conditions outlined, will consistently estimate the BATE and the CAB. To our knowledge, these estimators and their underlying assumptions have not been discussed in the context of binarization.

The rest of the paper will proceed as follows. In Section 2, we present the proof of equivalence between the BATE and the difference in average outcome of modified treatment policies that restrict the value of the exposure variable while preserving relative self-selection. We also introduce the CAB and discuss its relevance. Finally, we introduce regression adjustment-based, inverse probability weighting, and augmented inverse probability weighting estimators for the two estimands. Section 3 follows with a simulation study in which we implement the regression, inverse probability weighting, and augmented inverse probability weighting estimators for the BATE and CAB. In Section 4, we apply the method of binarization to evaluate the effect of proximity to oil and gas wells during gestation on the probability of giving birth to an underweight newborn in the state of California.  Lastly, in Section 5, we provide a discussion of the results.

\section{Methods}

\subsection{Setup and Notation}

Consider an observational study with baseline confounders $W \in \mathbb{R}^p$ with $p \in \mathbb{N}$,  a single exposure (possibly continuous) $A \in \mathbb{R}$, and a single outcome $Y \in \mathbb{R}$. Let the observed data be $O = (W, A, Y)$. Let supp$(A)$ and supp$(W)$ denote the support of the exposure and confounders, respectively. 

We denote the data generating distribution as $P_0$ for the observed data $O$. Assume that $O_1,...,O_n$ are $n$ i.i.d. draws of $O \sim P_0$. Let our statistical target parameter $\Psi: \mathcal{M} \rightarrow \mathbb{R}$ be a mapping from our statistical model to the parameter space (in this case, real line). Then, the estimand can be thought of as a function of different distributions where $\Psi(P_0)$ gives us our true estimand of interest.

Each set of observed values $\{O_1, ..., O_n\}$ can be fully described using an empirical probability distribution $P_n$ that puts point mass $\frac{1}{n}$ at each observation $O_i$. Each draw from $P_0$ generates a new $P_n$; therefore, $P_n$ is also a random distribution. We can now define our statistical estimator $\hat{\Psi}(P_n)$ as a function of our observed data that maps to the parameter space $\mathbb{R}$. Since $P_n$ is random, $\hat{\Psi}(P_n)$ is also random, meaning that each set of observed data $\{O_1, ..., O_n\}$ can correspond to a different value of the estimator. 

Let $\mathcal{A}$ be a subset of supp$(A)$ and $\mathcal{A}^{\mathsf{c}} = \text{supp}(A) / \mathcal{A}$ be the complement of $\mathcal{A}$. Consider a binary exposure variable $T$. An individual with a natural value of exposure $A=a$ is assigned a value of $T=1$ when $a \in \mathcal{A}$ and assigned $T=0$ when $a \in \mathcal{A}^{\mathsf{c}}$. In the traditional binarization setting, $\mathcal{A}$ is some interval on $\mathbb{R}$ (for example, $T=1$ when $A \geq a$, and $T=0$ when $A < a$, for some $a \in \mathbb{R}$). 

As is standard practice in causal inference literature, we will adopt the potential outcome framework \cite{rubin_estimating_1972}. The framework assumes that, for each individual and each permissible value of exposure $A=a$, there exists a counterfactual outcome $Y_a$ corresponding to the value that the outcome would have taken in a world in which the exposure were deterministically set to $a$. The observed outcome is $Y = Y_A$.

\subsection{Modified Treatment Policies}

In causal inference, modified treatment policies (MTPs) are used to consider counterfactual worlds in which the distribution of the exposure $A$ conditional on confounders $W$ is changed in some way \cite{haneuse_estimation_2013, diaz_munoz_population_2012}. The original measured exposure $A$ is called the ``observed" or ``status-quo" exposure, as it corresponds to the value of exposure an individual received in the observed world, prior to any intervention. We use $\tilde{A}$ to denote the post-intervention exposure, which can depend on $A$ and $W$. By considering changes in the underlying population distribution of the exposure, MTPs allow for stochastic interventions in addition to more traditional deterministic interventions. These types of interventions can therefore represent counterfactual worlds that are more realistic in modeling the type of behaviors that researchers may expect in a given population given a particular policy. One example of a MTP is a shift intervention, where researchers consider a policy where individuals with an observed exposure value $A=a$ would instead have a shifted exposure value $\tilde{A}=a+\delta$ for some $\delta \in \mathbb{R}$ \cite{diaz_munoz_population_2012}. Estimation strategies have been studied extensively for MTPs. For a comprehensive overview of modified treatment policies and nonparametric estimation strategies of causal effects, we refer the reader to Díaz et al. \cite{diaz_nonparametric_2023}.

\subsection{Equivalence of Binarized Average Treatment Effect and Average Treatment Effect Under Preservation of Relative Self-Selection}

Presume that we are analyzing the binarized data $(Y, T, W)$. A common practice is to regress $Y$ onto $T$ and $W$ to obtain an estimate of the regression function $E[Y|T,W]$ and then report the coefficient on $T$. This is a specific case of a more general plug-in estimation strategy where the estimand (binarized ATE; BATE) is given by 

\begin{equation}
    \psi_{\text{BATE}} = E[E[Y|T=1,W]] - E[E[Y|T=0,W]].
\end{equation}

\noindent In other words, we use the regression model to impute or predict \textit{both} ``potential outcomes'' for all subjects, then average across them all and take the difference. 

This estimand is represented in terms of the binarized exposure $T$. Our question is: \textbf{what causal contrast (if any) does this estimand represent in terms of the original exposure $A$?}

Consider the estimand $E[E[Y|T=1, W]]$. Using total expectation,
\begin{align*}
    E[E[Y|T=1, W]]
    &= E[E[Y|A \in \mathcal{A}, W]]\\
    &= E_W[E_{A|W,A\in \mathcal A}[E[Y|A, A\in \mathcal{A}, W]]]\\
    &= E_W[E_{A|W,A\in \mathcal A}[E[Y|A, W]]]\\
    & = E[\mu(\tilde A_1,W)],
\end{align*}

\noindent where we have defined $\mu(a,w) = E[Y|A=a,W=w]$ and a variable $\tilde A_1$ such that $\tilde A_1|W \sim A|W,A\in \mathcal A$ for every $W$. The density of $\tilde A_1|W$ follows directly from the conditioning: for $a \in \mathcal A$, we have $p_{A|W, A\in \mathcal{A}}(a,w) 
= p_{A|W}(a,w)/\pi_{\mathcal{A}}(w)$ where we have defined $\pi_\mathcal{A}(w) = Pr(A\in \mathcal{A}|W=w)$. Thus, extending to the support of $A$, the variable $\tilde A_1|W$ has density 

\begin{align*}
p_{\tilde A_1|W}(a,w) 
&= \frac{1_\mathcal{A}(a) p_{A|W}(a,w)}{\pi_{\mathcal{A}}(w)}.\\
\end{align*}

The distribution of $\tilde{A}_1$ preserves what we term \textit{relative self-selection} of exposure values within strata $W$ of the population (conditional on the exposures being within the desired region). Specifically, $\frac{p_{\tilde A_1|W}(a,w)}{p_{\tilde A_1|W}(a',w)} = \frac{p_{A|W}(a,w)}{p_{A|W}(a',w)}$ for any two exposure values $a, a' \in \mathcal A$. We can therefore think of $\tilde A_1$ as the unique exposure assignment mechanism $\tilde A$ for which $P(\tilde A \in \mathcal A)=1$ that also preserves the relative self-selection preferences expressed by $\frac{p_{A|W}(a,w)}{p_{A|W}(a',w)}$. 

To further illustrate the concept of relative self-selection, consider a simple case with three exposure $\{1,2,3\}$ and a population where the probabilities of receiving each exposure are $\left\{\frac{1}{4}, \frac{1}{4}, \frac{1}{2}\right\}$,  respectively. If we ``outlaw'' exposure 1 (i.e., threshold exposure so $\mathcal A = \{2,3\}$), then under treatment policy $\tilde A_1$, we would now have probability of $\frac{1}{3}$ of receiving exposure $A=2$ and probability of $\frac{2}{3}$ of receiving exposure $A=3$ (preserving the relative probabilities between the ``allowed" exposures within strata of the population). 

Preservation of relative self-selection is akin to Luce's definition of independence of irrelevant alternatives (IIA) \cite{ray_independence_1973}. In the definition of IIA, actors are typically considering choosing between only two options before the introduction of irrelevant alternatives. However, in the BATE case, we can compare the ratio of probability density of selecting one value of exposure $a$ relative to another $a'$ conditional on confounders given the choices present in the smaller set $\mathcal{A}$ or the choices present in the full support $\text{supp}(A)$. Under IIA, adding additional choices for exposure should not change the relative self-selection ratio. Therefore, interventions which act on individuals who follow IIA can be modeled by this MTP.

Given this exposition, we now define $\tilde A_0$ such that $\tilde A_0|W \sim A|A\in \mathcal{A}^{\mathsf{c}},W$. By identical arguments as above, $E[E[Y|T=0, W]] = E[\mu(\tilde A_0,W)]$. Therefore,

\begin{equation}
    \Psi_{\text{BATE}} = E[\mu(\tilde A_1,W)] - E[\mu(\tilde A_0,W)].
\end{equation}

\subsection{Identification and Causal Assumptions}
\label{identification}
The reason we have cast the binarized ATE in the form given above is because it has a very natural causal interpretation under standard causal assumptions. Specifically, assume:

\begin{enumerate}
    \item \textit{Positivity}: $0 < Pr(A \in \mathcal{A}|W=w) < 1$ for all $w \in \text{supp}(W)$.
    
    \item \textit{Conditional independence}: $Y_a \ind A  \,| \, W$ for all $a \in$ supp($A$).
\end{enumerate}

Consider an arbitrary modified exposure $\tilde A = \tilde A_1$ or $\tilde A = \tilde A_0$. Under these assumptions, we can write:

\begin{align*}
E[\mu(\tilde A, W)] 
&= E_{\tilde A,W}[E[Y|A=a, W=w]] \\
&= E_{\tilde A,W}[E[Y_A|A=a, W=w]]
\quad \text{(definition of $Y$)} \\
&= E_{\tilde A,W}[E[Y_a|A=a, W=w]] 
\quad \text{($Y_a=Y_A$ when $A=a$)} \\
&= E_{\tilde A,W}[E[Y_a|\tilde A=a, W=w]] 
\quad \text{(conditional independence)} \\
&= E_{\tilde A,W}[E[Y_{\tilde A}| W=w]] \\
&= E[Y_{\tilde A}].
\end{align*}

Note that the positivity assumption allows us to say that the expectations $E[Y|\tilde{A}_1, W]$ and $E[Y|\tilde{A}_0, W]$ will be well-defined. Moreover, the conditional independence between $Y_a$ and $A$ implies the same for $\tilde A_1$ and $\tilde A_0$. 

Therefore, finally:

\begin{equation}
    \psi_{\text{BATE}} = E[Y_{\tilde A_1}] -  E[Y_{\tilde A_0}].
\end{equation}

Under these assumptions, $\Psi_{BATE}$ is equivalent to the difference in expected outcome under two modified treatment policies that affect the underlying distribution of exposure in the population. Specifically, we can consider setting $T=1$ equivalent to enacting a policy that restricts $A \in \mathcal{A}$ while preserving relative self-selection, and setting $T=0$ equivalent to enacting a policy that restricts $A \in \Acomp$ while preserving relative self-selection. This interpretation is completely transparent and explicitly defines an intervention, negating concerns about ``different versions of treatment.'' Given the (standard) causal identification assumptions, this parameter is a perfectly reasonable causal contrast to use when estimating the effects of policies\textit{ under which persons would have exposure probabilities in the thresholded region proportional to what they would have been without the thresholding}. In Section \ref{Discussion}, we give examples of policies where this assumption is sensible and when it is not.

\subsection{New Parameter}
In the previous section, we showed that the BATE is equivalent to the difference between two average counterfactual outcomes under two different modified treatment policies, under causal assumptions. However, when considering the impact of a policy, it seems most natural to consider the difference between a counterfactual outcome under one MTP and the outcome under the status quo with no policy, as suggested by Hubbard and van der Laan \cite{hubbard_population_2008}. 
For example, the estimand proposed by Haneuse and Rotnitzky is a difference that ``quantifies the average effect in the studied population of a policy that switches each received dose\dots to the modified dose" \cite{haneuse_estimation_2013}. As another example, when studying the effect of shift interventions, researchers often estimate the difference between the average outcome under the proposed shift intervention and the observed average outcome \cite{munoz_population_2012, nugent_demonstration_2023}. 

To this end, we propose a new estimand of interest, the causal attributable effect of binarizing (CAB). The CAB is akin to the causal attributable risk (CAR) of the MTPs described, generalized to non-binary outcomes. The CAR is a common estimand in public health used to examine the excess risk that results from a particular exposure \cite{hoffman_chapter_2019}. It requires no data beyond what is used to calculate the BATE. The CAB is defined as the difference between the expected outcome under one of the two policies and the expected outcome under the observed world:

$$\Psi_{CAB} = E[Y_{\tilde{A}_t}] - E[Y] \,\,\, \text{for } t \in \{0, 1\}.$$

Estimands of interest should correspond with the question researchers are interested in exploring. Researchers are usually interested in asking causal questions to help guide decisions when faced with a choice. For example, often in medicine, researchers are interested in knowing whether a particular treatment is effective in decreasing the probability of a bad outcome relative to no treatment. Therefore, the ATE is an appropriate choice, as it compares a counterfactual world where everyone in the target population receives the treatment versus a counterfactual world in which no one in the target population receives the treatment. In the case of testing a brand-new treatment, the second counterfactual world essentially represents the observed, status-quo world, which currently exists without the treatment. In the case of considering a binarized exposure variable, we argue that the relevant choices are not between the two counterfactual worlds under two MTPs, but rather between one counterfactual world and the status-quo world. Put in other words, the relevant question is, ``On average, how would the outcome change from the status-quo if we were to restrict the exposure values into a particular interval of allowable values?" $\Psi_{CAB}$ is the answer to this question. 

In addition, if considering implementing a particular threshold, estimating the BATE rather than the CAB could result in researchers overstating the effect of such a policy. For example, suppose a hospital is considering implementing a policy in which patients with a systolic blood pressure greater than or equal to 140 mmHg are put on a medication to reduce the blood pressure to under 140 mmHg, and doctors want to know how this would affect the overall risk of patients at the hospital experiencing cardiac arrest. Here, if researchers were to estimate and report the BATE of implementing such a threshold on blood pressure, researchers would be comparing average risk of cardiac arrests in a world where all patients have a systolic blood pressure greater than or equal to 140 mmHg to that in a world where all patients have a systolic blood pressure under 140 mmHg. It is easy to see how such an estimate could overstate the effect of implementing such a policy because the counterfactual comparison world in which everyone has blood pressure greater than or equal to 140 mmHg overstates the current disease burden. Researchers could instead calculate the CAB to compare the risk of cardiac arrest in a counterfactual world in which all patients have systolic blood pressure under 140 mmHg to that in the current world, which would more accurately reflect the reduction doctors could reasonably expect. 

In addition to answering a more relevant research question, $\Psi_{CAB}$ should be convenient for researchers to consider in scenarios where they might consider $\Psi_{BATE}$. The identification assumptions are slightly weaker for the CAB than the BATE. Specifically, the positivity assumption becomes a one-way positivity assumption, which will depend on the particular causal contrast of interest. However, the two-way positivity assumption implies the one-way; therefore, any data generating process that allows for the identification of the BATE also allows for identification of the CAB. Thus, data generated by processes that satisfied conditions allowing for $\Psi_{BATE}$ to be estimated are also data in which $\Psi_{CAB}$ may be estimated. In the following section, we describe four methods by which researchers may implement estimators for $\Psi_{CAB}$, including details for calculating standard errors in Appendix \ref{appendix_est}.

\subsection{Estimators}\label{estimators}

Researchers can estimate the BATE using standard estimation techniques for the ATE by using the indicator $T$ as the binary treatment indicator. In Appendix \ref{appendix_est}, we provide details for a regression-based adjustment estimator \cite{ding_first_2023} and an inverse probability weighting (IPW) estimator \cite{robins_estimation_1994}. The regression estimator results from modeling the outcome regression $\mu(t,w) = E[Y|T=t,W=w]$ using a linear model and then plugging this estimate into the estimand of interest. The IPW estimator weights observations based on estimates of the propensity score, $\pi_{\mathcal{A}}(w) = Pr(T=1|W=w)$. We also provide details for two doubly-robust estimators: the augmented inverse probability weighted (AIPW) estimator \cite{robins_estimation_1994} and the targeted maximum likelihood estimator (TMLE) \cite{van_der_laan_nonparametric_2023}. We additionally provide details for how to use these estimation techniques to estimate the CAB. Though very similar, the estimators for the parameter and the variance differ slightly.

Under certain conditions, the proposed estimators are consistent for the true estimands, meaning that as the sample $n$ grows to infinity, the estimators will converge on the correct population level values of the estimands. Specifically, the regression estimator will be consistent if the outcome regression $E[Y|T=t,W=w]$ for $t \in \{0,1\}, w \in \text{supp}(W)$ is correctly specified; for the model described in Appendix \ref{appendix_reg}, we describe in more detail when this will be true in Appendix \ref{appendix_consistency}. The IPW estimator will be consistent if researchers have a correctly specified model for the propensity score $\pi_\mathcal{A}(w) = Pr(T=1|W=w)$ for $w \in \text{supp}(W)$. As previously mentioned, the AIPW and TMLE estimators are doubly-robust. This means that the estimators will be consistent estimators if researchers are able to either consistently estimate the outcome regression $E[Y|T=t, W=w]$ or the propensity score $\pi_\mathcal{A}(w)$ for $t \in \{0,1\}, w \in \text{supp}(W)$.

The regression \cite{stefanski_calculus_2002},  IPW \cite{lunceford_stratification_2004}, AIPW \cite{robins_estimation_1994}, and TMLE \cite{van_der_laan_nonparametric_2023} estimators are all asymptotically normal under certain conditions, allowing researchers the ability to construct valid confidence intervals based on the estimated standard errors. In Appendix \ref{appendix_est}, we provide variance estimators for the proposed methods. It is important to note that, while the AIPW and TMLE estimators are doubly-robust, the influence curve based estimators presented in the Appendix for the variance presented are not doubly-robust - they are only consistent when both the outcome regression and the propensity score estimators are consistent \cite{van_der_laan_unified_2003}. In addition, if both the outcome regression $E[Y|T=t, W=w]$ and the propensity score $\pi_\mathcal{A}(w)$ are consistently estimated for $T \in \{0,1\}, w \in \text{supp}(W)$, then the AIPW and TMLE estimators are semi-parametrically efficient, meaning that they have the smallest variance as the sample size grows to infinity in settings where we assume no knowledge on the distributions of the observed variables. 

As noted above, the four estimators rely upon estimators of conditional expectations or conditional probabilities. Given that the MTPs defined above result in exposure distributions that depend on the distribution of the observed $A$, one may assume that one must first estimate the density of $A$. However, the equivalence result shows that the only relevant piece of the distribution of $A$ that researchers need to estimate $\Psi_{BATE}$ and $\Psi_{CAB}$ is the distribution of $T$, the binarized exposure. More concretely, if researchers only had access to the data $(W, T, Y)$ for individuals and no other information about the actual value of exposure $A$, they would still be able to consistently estimate $\Psi_{BATE}$ and $\Psi_{CAB}$.  Our result also shows that, to estimate $\Psi_{BATE}$, one can use any standard ATE estimation technique plugging in $T$ as the binary exposure variable. However, the estimators for $\Psi_{CAB}$ take a different form, which can be found in Appendix \ref{appendix_est}.

\section{Simulation Study}
\subsection{Data Generating Processes}

To illustrate the equivalence between the binarized ATE and difference in expected outcomes under the proposed MTPs, we performed a simulation study. The data generating process (DGP) is as follows:

\begin{align*}
    W &\sim \text{ Bern}(p = 0.5)\\ 
    A|W &\sim \mathcal{N}(5+2*W, 1)\\
    Y|A, W &\sim A^3 + \text{sin}(A) + 100*W + \mathcal{N}(0,1)\\
    T | A &= \begin{cases}
        1 \text{ when } A \geq 6\\
        0 \text{ when } A < 6\\
    \end{cases}
\end{align*}

The estimands of interest are the binarized ATE and the causal attributable effect of binarization when implementing a cut-off at an exposure value of $A \geq 6$. Here, we define the CAB as $\Psi_{CAB} = E[Y_{\tilde{A}_1}] - E[Y]$, which contrasts the average outcome under the MTP which restricts values $A \geq 6$ and preserves relative self-selection with the average outcome in the observed world. Under the data generating process, the true parameter values are $\Psi_{BATE} = 201.806$ and $\Psi_{CAB} = 90.872$. 

Over 5000 simulations, we drew samples of size $n = 150, 300, 500$ from the data generating process and calculated the estimates $\widehat{\Psi}_{BATE}$ and $\widehat{\Psi}_{CAB}$ using the regression, IPW, and AIPW estimators discussed detailed in Appendix \ref{appendix_est}. The regression estimator is consistent for this DGP because $E[Y|T,W]$ is linear in $T$ and $W$, as $W$ is binary. For a more general discussion of when the regression estimator will be consistent, see Appendix \ref{appendix_reg}. The estimate $\hat{\pi}_\mathcal{A}(W)$ necessary for implementing the IPW estimator was calculated using logistic regression with $W$ as the predictor and $T$ as the binary outcome, which is consistent for the propensity score for this DGP. Finally, the AIPW estimator is doubly robust. Therefore, all three estimators are consistent. We also estimated the standard errors of the estimates using the methods described in Section \ref{estimators}. Finally, we present the bias of the estimates as well as the standard error of the estimators over the simulations to evaluate the performance of the estimators.

\subsection{Visualizing the MTPs}

\begin{figure}[H]
    \centering
    \includegraphics[width=0.8\textwidth]{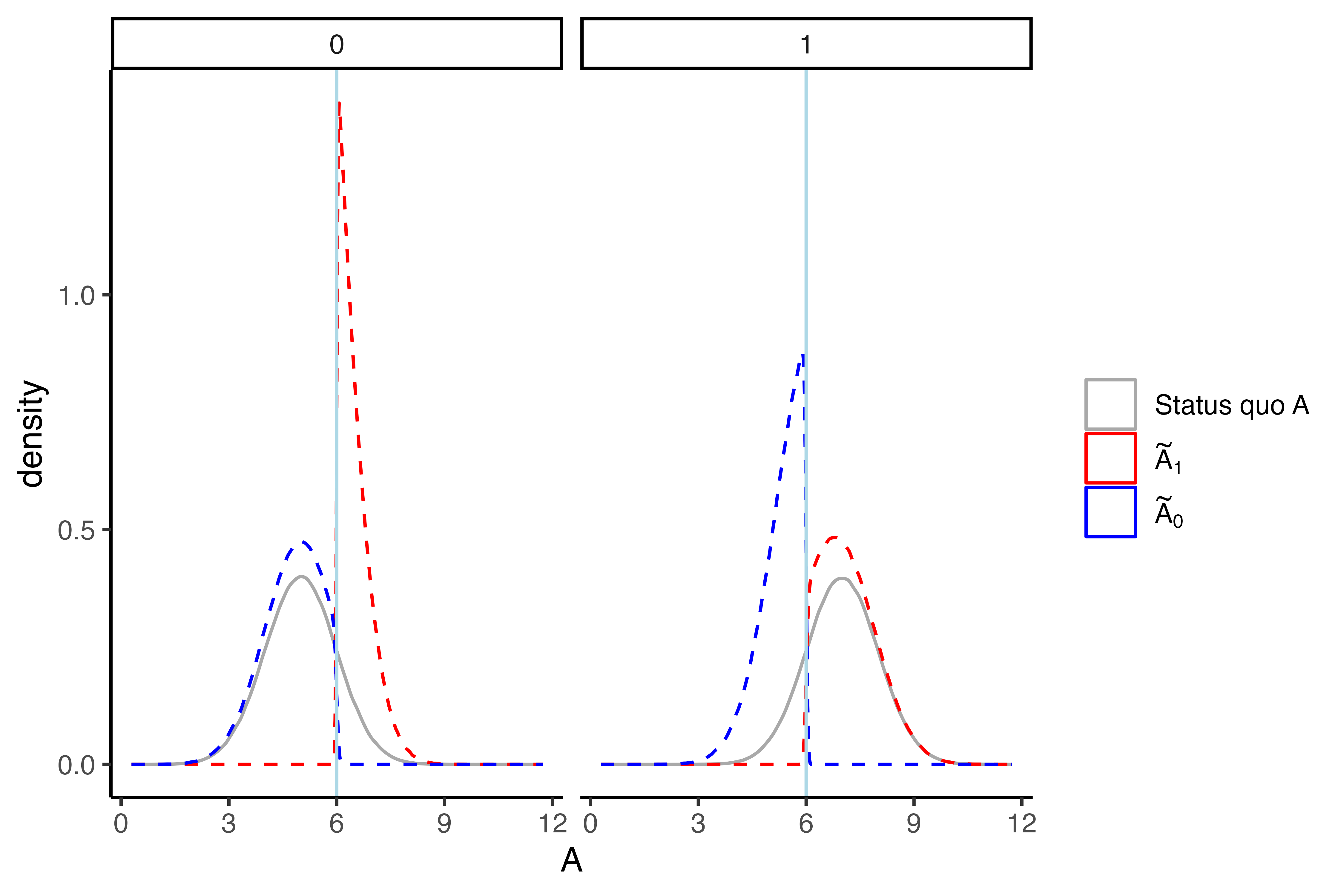}
    \caption{Densities of exposures $A$, $\tilde{A}_1$, and $\tilde{A}_0$ by different values of $W$. The graph on the left shows the densities when $W=0$ and graph on the right shows the densities when $W=1$. The green line represents the cut-off value of $A=6$. The gray line is the density of the observed $A$. The red dotted line is the density of $\tilde{A}_1$. The blue dotted line is the density of $\tilde{A}_0$.}
    \label{fig:sim_viz}
\end{figure}

Figure \ref{fig:sim_viz} illustrates the resulting densities after implementing the cut-off for the two values of the confounder $W$. The green line represents the cut-off value of 6. The gray line is the density of $A$ in the observed world, the red dotted line is the density of $\tilde{A}_1$, and the blue dotted line is the density of $\tilde{A}_0$. As illustrated by this figure, the densities $\tilde{A}_1$ and $\tilde{A}_0$ differ depending on the value of the confounder $W$. This is generally true of our result; the resulting distributions of exposure after binarizing can depend both on the distribution of the observed exposure as well as the pre-exposure confounders. 

Both $\tilde{A}_1$ and $\tilde{A}_0$ have density equal to 0 for values of exposure that are not within the allowed range ($A < 6$ for $\tilde{A}_1$ and $A \geq 6$ for $\tilde{A}_0$). It is also evident that both preserve relative self-selection for the values of exposure within the allowed range. The non-zero portions of the MTP distributions of exposure are simply scaled versions of the density of the observed $A$.

\subsection{Simulation Results}
\begin{landscape}

\begin{table}[H]
\centering
\begin{tabular}{|l|cccc|cccc|cccc|}
\hline
\multicolumn{1}{|c|}{}  & \multicolumn{4}{c|}{Regression Estimator}                                                             & \multicolumn{4}{c|}{IPW Estimator}                                                                    & \multicolumn{4}{c|}{AIPW Estimator}                                                                   \\ \hline
\multicolumn{1}{|c|}{n} & \multicolumn{1}{c|}{Estimate} & \multicolumn{1}{c|}{Bias (\%)} & \multicolumn{1}{c|}{Est SD} & Sim SE & \multicolumn{1}{c|}{Estimate} & \multicolumn{1}{c|}{Bias (\%)} & \multicolumn{1}{c|}{Est SE} & Sim SE & \multicolumn{1}{c|}{Estimate} & \multicolumn{1}{c|}{Bias (\%)} & \multicolumn{1}{c|}{Est SD} & Sim SE \\ \hline
150                     & \multicolumn{1}{c|}{202.121}  & \multicolumn{1}{c|}{-0.00149}  & \multicolumn{1}{c|}{13.671} & 14.718 & \multicolumn{1}{c|}{202.672}  & \multicolumn{1}{c|}{0.00123}   & \multicolumn{1}{c|}{14.684} & 14.688 & \multicolumn{1}{c|}{202.672}  & \multicolumn{1}{c|}{0.00123}   & \multicolumn{1}{c|}{14.419} & 14.457 \\ \hline
300                     & \multicolumn{1}{c|}{202.160}  & \multicolumn{1}{c|}{-0.00130}  & \multicolumn{1}{c|}{9.803}  & 10.179 & \multicolumn{1}{c|}{202.112}  & \multicolumn{1}{c|}{-0.00154}  & \multicolumn{1}{c|}{10.119} & 10.275 & \multicolumn{1}{c|}{202.112}  & \multicolumn{1}{c|}{-0.00154}  & \multicolumn{1}{c|}{10.118} & 9.866  \\ \hline
500                     & \multicolumn{1}{c|}{202.306}  & \multicolumn{1}{c|}{-0.00058}  & \multicolumn{1}{c|}{7.634}  & 7.781  & \multicolumn{1}{c|}{202.324}  & \multicolumn{1}{c|}{-0.00048}  & \multicolumn{1}{c|}{7.841}  & 7.860  & \multicolumn{1}{c|}{202.324}  & \multicolumn{1}{c|}{-0.00048}  & \multicolumn{1}{c|}{7.841}  & 7.860  \\ \hline
\end{tabular}
\caption{Simulation results for estimating $\Psi_{BATE}$. The true value of the estimand is 201.806. The ``Estimate'' and ``Bias'' columns contains the average of the BATE estimates and bias over the 5000 replicates, respectively. ``Est SD'' column contains the average of the estimated standard deviations over the 5000 replicates. Finally, the ``Sim SE'' column contains the standard error of the estimates over the 5000 replicates.}
\label{tab:res_bate}
\end{table}

\begin{table}[H]
\centering
\begin{tabular}{|l|cccc|cccc|cccc|}
\hline
\multicolumn{1}{|c|}{}  & \multicolumn{4}{c|}{Regression Estimator}                                                             & \multicolumn{4}{c|}{IPW Estimator}                                                                    & \multicolumn{4}{c|}{AIPW Estimator}                                                                   \\ \hline
\multicolumn{1}{|c|}{n} & \multicolumn{1}{c|}{Estimate} & \multicolumn{1}{c|}{Bias (\%)} & \multicolumn{1}{c|}{Est SD} & Sim SE & \multicolumn{1}{c|}{Estimate} & \multicolumn{1}{c|}{Bias (\%)} & \multicolumn{1}{c|}{Est SD} & Sim SE & \multicolumn{1}{c|}{Estimate} & \multicolumn{1}{c|}{Bias (\%)} & \multicolumn{1}{c|}{Est SD} & Sim SE \\ \hline
150                     & \multicolumn{1}{c|}{90.083}   & \multicolumn{1}{c|}{0.00050}   & \multicolumn{1}{c|}{12.273} & 11.757 & \multicolumn{1}{c|}{90.186}   & \multicolumn{1}{c|}{0.00164}   & \multicolumn{1}{c|}{11.464} & 11.421 & \multicolumn{1}{c|}{90.186}   & \multicolumn{1}{c|}{0.00164}   & \multicolumn{1}{c|}{11.317} & 11.421 \\ \hline
300                     & \multicolumn{1}{c|}{89.863}   & \multicolumn{1}{c|}{-0.00194}  & \multicolumn{1}{c|}{8.743}  & 7.845  & \multicolumn{1}{c|}{89.979}   & \multicolumn{1}{c|}{-0.00065}  & \multicolumn{1}{c|}{7.925}  & 8.050  & \multicolumn{1}{c|}{89.979}   & \multicolumn{1}{c|}{-0.00065}  & \multicolumn{1}{c|}{7.925}  & 8.050  \\ \hline
500                     & \multicolumn{1}{c|}{89.867}   & \multicolumn{1}{c|}{-0.00191}  & \multicolumn{1}{c|}{6.795}  & 6.038  & \multicolumn{1}{c|}{90.107}   & \multicolumn{1}{c|}{0.00077}   & \multicolumn{1}{c|}{6.150}  & 6.183  & \multicolumn{1}{c|}{90.107}   & \multicolumn{1}{c|}{0.00077}   & \multicolumn{1}{c|}{6.150}  & 6.183  \\ \hline
\end{tabular}
\caption{Simulation results for estimating $\Psi_{CAB}$. The true value of the estimand is 90.872. The ``Estimate'' and ``Bias'' columns contains the average of the CAB estimates and bias over the 5000 replicates, respectively. ``Est SD'' column contains the average of the estimated standard deviations over the 5000 replicates. Finally, the ``Sim SE'' column contains the standard error of the estimates over the 5000 replicates.}
\label{tab:res_CAB}
\end{table}

\end{landscape}

Using the methods described in Appendix \ref{appendix_est}, we calculated estimates for the target parameters and the standard errors of the estimates. What bears repeating is that \textit{the only information the estimators require about the exposure $A$ is the value of the binarized exposure $T$.} One may assume that, in order to estimate the expected outcomes under the MTPs, one must first estimate the full densities of $\tilde{A}_1$ and $\tilde{A}_0$. However, through the equivalence with the binarized ATE, the relevant quantity one must estimate is instead the conditional probability $\pi_{\mathcal{A}}(w)= Pr(T = 1 | W=w) = Pr(A \geq 6 | W=w)$ for $w\in \{0,1\}$, which is easier to estimate. 

As the results in tables \ref{tab:res_bate} and \ref{tab:res_CAB} show, the regression, IPW, and AIPW estimators for $\Psi_{CAB}$ and $\Psi_{BATE}$ have almost negligible bias, even for the smaller sample size of $n=150$. When estimating both $\Psi_{CAB}$ and $\Psi_{BATE}$, the regression, IPW, and AIPW estimators have relatively similar standard errors, and thus would result in relatively similar normal-based confidence intervals. 

\section{Application: Oil/Gas Well Exposure and Birth Outcomes in California, 2007--2015}

To demonstrate how to think about binarization in an applied setting, we revisit the questions analyzed by Tran et al. investigating the effect of residential proximity of oil/gas wells during pregnancy on birth outcomes in the state of California from 2006--2015 \cite{tran_et_al}. The authors found a significant association between high production volume from oil and gas wells and small gestational age (SGA) in rural and urban areas. In response to``a growing body of research shows direct health impacts from proximity to oil extraction," California Senate passed a law creating Health Protection Zones from oil and gas wells starting January 1, 2023 \cite{ca_sb_1137}. Health Protection Zones are 3,200 feet ($\approx 1$ kilometer) buffers surrounding homes, schools, hospitals, and other sensitive locations. To evaluate the effect of such a law on the probability of SGA in newborns in California based on the 2007-2015 data, we considered a binarized exposure, where infants born to households within 1km of oil and gas wells are considered ``untreated," and infants born to households outside of 1km of an active oil/gas well are considered ``treated."

Our population of interest is infants born in California between 2007 and 2015 where the residential address of the pregnant parent was within 10km of an active oil/gas well during the period of gestation, as in Tran et al - we followed the same exclusion criteria as in Figure 1.  We consider the same infant and maternal confounding factors as in Tran et al.; specifically, we adjust for infant sex, month (categorical) of birth, year of birth, maternal age in years, maternal race/ethnicity, maternal educational attainment, Kotelchuk index of prenatal care, and parity. All study protocols were approved by the Institutional Review Board of the CA Department of Public Health (\#13-05-z). 

Let $\mathbf{W}$ be the vector of covariates, $D$ be the distance in meters to the closest well, and $Y$ be the indicator of low birth weight (defined as a birthweight of $< 2500g$ \cite{tran_et_al}). Some parents may live within 10km of multiple wells --- for the purposes of this analysis, we chose the distance to the closest well as our exposure of interest. We then binarize this exposure at 1km (assigning individuals with a distance $D \leq 1$km to $T=0$ and a distance $D > 1$km $T=1$. Our estimands of interest are $\Psi_{BATE} = E[E[Y|T=1, \mathbf{W}]]  - E[E[Y|T=0, \mathbf{W}]]$ and $\Psi_{CAB} = E[E[Y|T=1, \mathbf{W}]]  - E[E[Y| \mathbf{W}]]$. 

To estimate these quantities, we use the AIPW estimator presented in Appendix \ref{appendix_AIPW}. We estimate the outcome regression $E[Y|T, \mathbf{W}]$ using a logistic regression estimator: $\widehat{m}(T,\mathbf{W}) = \text{expit}( \widehat{\beta_0} + T\widehat{\beta_T} + \mathbf{W}\widehat{\mathbf{\beta}}_{\mathbf{W}} + T\mathbf{W}\widehat{\mathbf{\beta}}_{interact}$). To estimate the propensity score ($Pr(D > 1\text{km} | \mathbf{W})$), we use a logistic regression estimator: $\widehat{\pi_T}(w) = \text{expit}\left( \widehat{\beta}_0 + \mathbf{W}\widehat{\beta}_\mathbf{W}\right)$. If either of these estimators are consistent for the true function, then the estimates will be consistent. To estimate the standard deviation, we implement the nonparametric bootstrap. The results are presented in Table \ref{tab:res_births}. 

\begin{table}[H]
\centering
\begin{tabular}{l|c|c|}
\cline{2-3}
                                                    & Estimate & SD          \\ \hline
\multicolumn{1}{|l|}{$\Psi_{BATE}$} & 0.205\%  & (0.110\%)   \\ \hline
\multicolumn{1}{|l|}
{$\Psi_{CAB}$}  & 0.0080\% & (0.0049\%) \\ \hline
\end{tabular}

\caption{Estimates for $\Psi_{BATE}$ and $\Psi_{CAB}$ and the corresponding standard deviations when binarizing distance to closest active oil/gas well at 1km.}
\label{tab:res_births}
\end{table}

The key identifying assumption underlying our analysis is that, conditional on observed covariates, the distribution of distances to the nearest well among those who would be ``treated'' under the law (i.e., living more than 1 km from an active well) is proportional to the distribution observed in the current world. In our notation, 
$$
P(D|T=1, \mathbf{W}) \propto P(D|\mathbf{W}).
$$

\begin{figure}[H]
    \centering
    \subfigure{\includegraphics[width=0.4\textwidth]{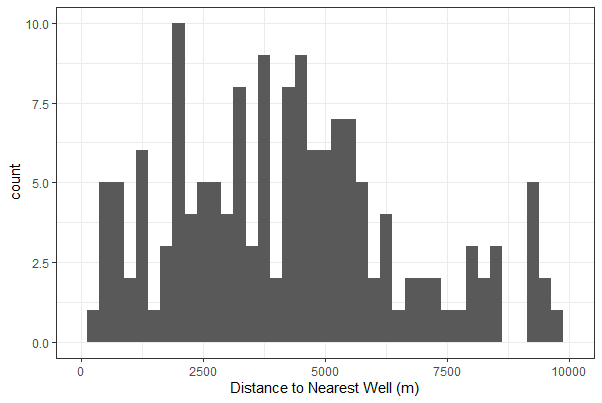}}
    \subfigure{\includegraphics[width=0.4\textwidth]{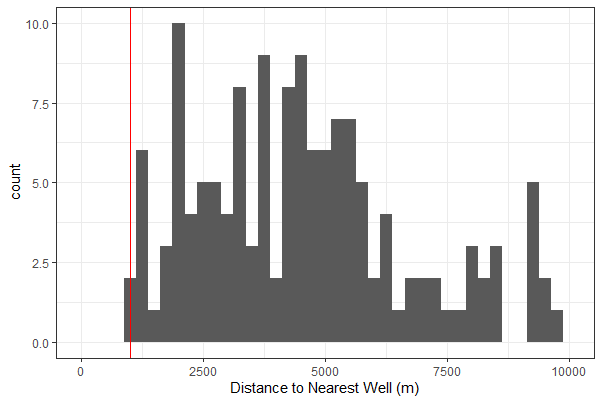}}
    
    \caption{Histograms for the observed distance to the nearest active well to households with pregnant Hispanic people, aged 25–29, less than high school educated with a Kotelchuk index of adequate and nulliparous, who gave birth in September 2010 to male babies. The histogram on the left is the observed distribution. The histogram on the right is the observed histogram after imposing a cut-off at 1 km; this is approximately what we assume the distribution of distance to nearest active well would look like after implementing the Health Protection Zones.}
    \label{fig:hist}
\end{figure}

Intuitively, this means that within any stratum of pregnant people with the same observed covariates (e.g., Hispanic, aged 25–29, less than high school educated with a Kotelchuk index of adequate and nulliparous, giving birth in September 2010 to male babies), the histogram of distances to the nearest well in the post-law world should look like the observed histogram, but truncated at 1 km. A concrete way to visualize this is to imagine taking the empirical distribution of distances for a covariate subgroup and slicing it at the 1 km threshold, as in Figure \ref{fig:hist}. The remaining portion approximates what we expect under the intervention. While we only observe finite samples, the assumption is about the population distribution and must hold for all covariate strata. If this were not the case, the effect we estimate via a binarized analysis would not correspond to the real-world effect of closing or relocating wells within a 1km radius of populated areas. 

It may help to consider what this means in terms of the California law. When the law is enacted, households do not relocate --- instead, the wells themselves are removed (or relocated beyond the buffer). Thus, the intervention acts by eliminating wells within 1 km of residences and perhaps relocating them outside of 1 km. Under the binarization assumption, the distances to the nearest well beyond 1 km remain unchanged. In reality, however, removing wells near households will often increase the distances to the nearest active well for other households, since the nearest active well may now be farther away. Therefore, the assumption is not literally true, though the degree of violation depends on the density and spatial configuration of wells. When wells are sparse, the shift could be substantial; when wells are dense, the effect may be negligible.

Finally, it is important to emphasize that the binarization assumption is not the only requirement for causal validity. A causal interpretation of this analysis still depends on the assumptions in Section \ref{identification} holding for our analysis. For example, we must have no unmeasured confounding. Despite controlling for maternal demographics, education, prenatal care, and birth timing, unmeasured socioeconomic or environmental factors (some of which were included in Tran et al.) could bias results. In addition, our parametric modeling choices introduce assumptions. Logistic regressions for outcome regressions and propensity scores are unlikely to be correctly specified in this context, which can induce bias. More flexible machine learning methods with cross-validation could mitigate these concerns, but pursuing that is beyond the scope of this paper. We therefore interpret our results as illustrative of how binarization can be applied, rather than definitive evidence of health effects.

Assuming the assumptions outlined above hold, $\Psi_{BATE}$ is equivalent to the difference in probability of low birth weight under the intervention that restricts active oil/gas wells to outside of 1km of homes and the intervention that restricts active oil/gas wells to within 1km of homes. Under the same assumptions, $\Psi_{CAB}$ is equivalent to the difference in probability of low birth weight under the intervention that restricts active oil/gas wells to outside of 1km of homes and the observed world. The two estimates are around $0.205\%$ and $0.0080\%$, respectively.

The results underscore the important distinction between the BATE and CAB, as the BATE can overestimate the effect of an intervention. If researchers had only considered the BATE, the estimate would suggest a $0.205\%$ increase in the probability of a baby being born with a low birthweight. However, considering the CAB, which compares the post-intervention world with the status quo world, we find only a $0.0080\%$ increase in the probability. 

It may seem surprising that the BATE estimate here is positive, implying that a policy moving all wells outside of 1 km of households would increase the probability of low birth weight compared to a policy moving all wells within 1 km of households. There are two reasons our binarized estimand may be understating the protective effect of the policy. The first is that, as explained above, if the health harms of proximity decay smoothly with distance, then removing nearby wells should reduce exposure more than what is captured by a strict 1 km cut-off. Put differently, the real law is expected to do at least as well as, and likely better than, the scenario we analyze. On the other hand, if effects beyond 1 km are truly negligible, then the difference is immaterial, and the binarization assumption becomes a reasonable approximation for causal inference. The second is the exposure we consider, distance to closest well, may not fully capture the health harms of allowing active wells to be built within 1 km of households. For example, some individuals may live within 1 km of several wells, and some active wells may have higher levels of production than others. For example, the authors of the Tran et al. paper found that households exposed to high production volume had an average of 32 active wells within 1 km of the household. Thus, considering only the closest well may not capture the full health harms pregnant people are exposed to prior to the buffer zone being implemented. The fact that the estimates are very close to 0 lends further credence to this idea, suggesting that the distance to closest well may not be a good proxy for the amount of harm an individual may experience living within 1km of active wells.

\label{Discussion}
\section{Discussion}

In this paper, we show the equivalence between the binarized ATE and the difference between the average counterfactual outcomes under two MTPs that impose the relevant cut-off defined by the binarization and preserve relative self-selection. Relative self-selection is the ratio of the probability density of one value of exposure to another. Through this equivalence, we affirm that the binarized ATE is a valid causal estimand and is identified under standard causal identification assumptions, addressing many of the common concerns methods researchers have posited about such coarsening of a continuous exposure variable. 

We also introduce a new estimand, the causal attributable effect of binarization (CAB), which contrasts the expected outcome under one of the MTPs associated with binarization with the expected outcome in the  observed world. Identification of this estimand requires weaker assumptions, and these assumptions are implied by the assumptions required to identify the BATE --- thus, a scenario in which researchers could estimate the BATE is also a scenario in which they could estimate the CAB. We argue that this estimand answers a more relevant causal question as it compares the counterfactual world under one MTP (the ``treated" counterfactual) to the observed world (which is akin to a ``control" counterfactual). As this estimand cannot be calculated using standard methods developed for the ATE, we also provide details for how to implement four different estimators of the causal attributable effect of binarization as well as how to estimate their standard errors. 

We also conducted a simulation study to show analytically that the results hold using the estimators that we proposed. Specifically, using only the binarized variable $T$ based on the cut-off and no other information about the observed exposure distribution of $A$, we were able to provide consistent estimates for both the BATE and the CAB. This result may be surprising, as typically estimating the effects of MTPs requires estimating the density of $A$. For example, this is the case in the shift intervention considered by Díaz and van der Laan \cite{munoz_population_2012}. Therefore, the two MTPs discussed which impose a cut-off and preserve relative self-selection lend themselves to much easier estimation from a practioner's point of view, which perhaps explains part of the reason why the method is so prevalent in the literature across disciplines. In fact, it has been shown that for all MTPs which impose a cut-off and restrict exposures to a certain region of the support, the causal effects for the unique MTP which preserves relative self-selection are indeed the easiest to estimate from a theoretical view point \cite{van_der_laan_nonparametric_2023}. Finally, we consider an applied example, applying binarization to births in California to evaluate a 1 km buffer zones between active oil/gas wells and homes.

Although the BATE and CAB may be easier to estimate, they may not be relevant causal quantities depending on the specific research context. Ultimately, whether binarization should be employed as detailed in this paper depends on the expected behavior of individuals after implementing the intervention of interest. For example, suppose researchers were interested in estimating the effect of a policy that limits the amount of energy households  use by fining households that consume more energy than a certain threshold. In this case, the assumption of preservation of relative self-selection is not very plausible, as one could reason that households above the threshold would change their energy usage to just barely meet the new limits, as there is no incentive to reduce energy consumption any further. Conversely, households that already use less energy than the threshold will likely make no change to their energy use. Therefore, the resulting distribution of household energy usage would have a ``piling up" of households at the imposed threshold with the distribution below the threshold looking the same as that in the observed world. Therefore, in this case, it would likely not make sense to binarize household energy usage based on the proposed cut-off. 

On the other hand, suppose researchers were interested in estimating the effect of the distance an individual lives from a manufacturing plant. Specifically, the researchers are interested in estimating the effect of a proposed policy which would give individuals who currently live within a specified cut-off distance of the plant a housing voucher which they can use to relocate throughout the city. In this case, it might be more reasonable that preservation of relative self-selection would hold after implementing the policy. Deciding where to live is a multifaceted decision based on individual needs that may depend on different factors such as access to public transportation, quality of local schools, and proximity to grocery stores, among other factors. The individuals who chose to live close to the factory may have done so based on different needs; therefore, it seems plausible that these individuals would relocate evenly across the other populated areas of the city if given the economic assistance to do so. 

These are simply two examples of policies researchers may be interested in. Ultimately, the decision to binarize should depend on domain-specific knowledge and predictions regarding the effect of the relevant policy on the distribution of the exposure variable in the population. Unfortunately, before the policy is implemented, this is a completely untestable assumption. However, researchers can still make plausible educated guesses and choose a causal estimand of interest accordingly.

\section*{Acknowledgments}

The authors thank Rachel Morello-Frosch and David J.X. González for their invaluable support throughout this project. We are especially grateful for their generosity in providing us with access to the data used in our paper. The authors also thank Laura Balzer for her comments on an early draft.

\section*{Funding information}
This material is based upon work supported by the National Science Foundation Graduate Research Fellowship Program under Grant No. DGE 2146752. Any opinions, findings, and conclusions or recommendations expressed in this material are those of the author(s) and do not necessarily reflect the views of the National Science Foundation. The research was partially supported by funds from the NIEHS Superfund Research Program P42ES004705.

\section*{Author contributions}
All authors have accepted responsibility for the entire content of this manuscript and approved its submission.

\section*{Conflict of interest}
Authors state no conflict of interest.

\section*{Ethical approval}

The research related to human use has been complied with all the relevant national regulations, institutional policies and in accordance the tenets of the Helsinki Declaration, and has been approved by the authors' institutional review board or equivalent committee.

\section*{Data availability statement}

The code generating the simulation study data and analyzing the simulation and applied data are available at  \url{https://github.com/kaitlynjlee/bridging_binarization}. The births data that support the findings of this study are available from the California Department of Public Health, but restrictions apply to the availability of these data, which are governed under an IRB protocol to protect the confidentiality of participants. The well data are a subset of the California Geologic Energy Management Division's (CalGEM) All Wells dataset, available at \url{https://www.conservation.ca.gov/calgem/maps/Pages/GISMapping2.aspx}.

\bibliography{main}

@article{wu_causal_2019,
	title = {Causal inference in the context of an error prone exposure: {Air} pollution and mortality},
	volume = {13},
	issn = {1932-6157, 1941-7330},
	shorttitle = {Causal inference in the context of an error prone exposure},
	url = {https://projecteuclid.org/journals/annals-of-applied-statistics/volume-13/issue-1/Causal-inference-in-the-context-of-an-error-prone-exposure/10.1214/18-AOAS1206.full},
	doi = {10.1214/18-AOAS1206},
	number = {1},
	urldate = {2023-07-13},
	journal = {The Annals of Applied Statistics},
	author = {Wu, Xiao and Braun, Danielle and Kioumourtzoglou, Marianthi-Anna and Choirat, Christine and Di, Qian and Dominici, Francesca},
	month = mar,
	year = {2019},
	note = {Publisher: Institute of Mathematical Statistics},
	pages = {520--547},
}

@misc{rose_recoding_2023,
	title = {On {Recoding} {Ordered} {Treatments} as {Binary} {Indicators}},
	url = {http://arxiv.org/abs/2111.12258},
	urldate = {2023-07-13},
	publisher = {arXiv},
	author = {Rose, Evan K. and Shem-Tov, Yotam},
	month = jul,
	year = {2023},
	note = {arXiv:2111.12258 [econ, stat]},
}

@article{carneiro_average_2017,
	title = {Average and {Marginal} {Returns} to {Upper} {Secondary} {Schooling} in {Indonesia}},
	volume = {32},
	copyright = {Copyright © 2016 John Wiley \& Sons, Ltd.},
	issn = {1099-1255},
	url = {https://onlinelibrary.wiley.com/doi/abs/10.1002/jae.2523},
	doi = {10.1002/jae.2523},
	language = {en},
	number = {1},
	urldate = {2023-07-13},
	journal = {Journal of Applied Econometrics},
	author = {Carneiro, Pedro and Lokshin, Michael and Umapathi, Nithin},
	year = {2017},
	note = {\_eprint: https://onlinelibrary.wiley.com/doi/pdf/10.1002/jae.2523},
	pages = {16--36},
}

@article{kurtz_does_2022,
	title = {Does {Free}-{Market} {Reform} {Induce} {Protest}? {Selection}, {Post}-{Treatment} {Bias}, and {Depoliticization}},
	volume = {52},
	issn = {0007-1234, 1469-2112},
	shorttitle = {Does {Free}-{Market} {Reform} {Induce} {Protest}?},
	url = {https://www.cambridge.org/core/journals/british-journal-of-political-science/article/does-freemarket-reform-induce-protest-selection-posttreatment-bias-and-depoliticization/69016C9E4276CCB22F3A0CA493FD7E68},
	doi = {10.1017/S0007123420000605},
	language = {en},
	number = {2},
	urldate = {2023-07-17},
	journal = {British Journal of Political Science},
	author = {Kurtz, Marcus J. and Lauretig, Adam},
	month = apr,
	year = {2022},
	note = {Publisher: Cambridge University Press},
	pages = {968--976},
}

@article{hu_diet_2001,
	title = {Diet, {Lifestyle}, and the {Risk} of {Type} 2 {Diabetes} {Mellitus} in {Women}},
	volume = {345},
	issn = {0028-4793},
	url = {https://www-nejm-org.libproxy.berkeley.edu/doi/10.1056/NEJMoa010492},
	doi = {10.1056/NEJMoa010492},
	number = {11},
	urldate = {2023-07-17},
	journal = {New England Journal of Medicine},
	author = {Hu, Frank B. and Manson, JoAnn E. and Stampfer, Meir J. and Colditz, Graham and Liu, Simin and Solomon, Caren G. and Willett, Walter C.},
	month = sep,
	year = {2001},
	note = {Publisher: Massachusetts Medical Society},
	pages = {790--797},
}

@article{bennette_against_2012,
	title = {Against Quantiles: Categorization of Continuous Variables in Epidemiologic Research, and Its Discontents},
	volume = {12},
	issn = {1471-2288},
	shorttitle = {Against quantiles},
	url = {https://doi.org/10.1186/1471-2288-12-21},
	doi = {10.1186/1471-2288-12-21},
	number = {1},
	urldate = {2023-07-17},
	journal = {BMC Medical Research Methodology},
	author = {Bennette, Caroline and Vickers, Andrew},
	month = feb,
	year = {2012},
	pages = {21},
}

@article{vanderweele_causal_2013,
	title = {Causal {Inference} {Under} {Multiple} {Versions} of {Treatment}},
	volume = {1},
	issn = {2193-3677},
	url = {https://www.ncbi.nlm.nih.gov/pmc/articles/PMC4219328/},
	doi = {10.1515/jci-2012-0002},
	number = {1},
	urldate = {2023-07-20},
	journal = {Journal of causal inference},
	author = {VanderWeele, Tyler J. and Hernán, Miguel A.},
	month = may,
	year = {2013},
	pmid = {25379365},
	pmcid = {PMC4219328},
	pages = {1--20},
}

@article{vanderweele_inference_2011,
	title = {Inference for causal interactions for continuous exposures under dichotomization},
	volume = {67},
	issn = {0006-341X},
	url = {https://www.ncbi.nlm.nih.gov/pmc/articles/PMC3178663/},
	doi = {10.1111/j.1541-0420.2011.01629.x},
	number = {4},
	urldate = {2023-07-20},
	journal = {Biometrics},
	author = {VanderWeele, Tyler J. and Chen, Yu and Ahsan, Habibul},
	month = dec,
	year = {2011},
	pmid = {21689079},
	pmcid = {PMC3178663},
	pages = {1414--1421},
}

@article{hubbard_population_2008,
	title = {Population intervention models in causal inference},
	volume = {95},
	issn = {0006-3444},
	doi = {10.1093/biomet/asm097},
	language = {eng},
	number = {1},
	journal = {Biometrika},
	author = {Hubbard, Alan E. and van der Laan, Mark J.},
	year = {2008},
	pmid = {18629347},
	pmcid = {PMC2464276},
	pages = {35--47},
}

@article{haneuse_estimation_2013,
	title = {Estimation of the effect of interventions that modify the received treatment},
	volume = {32},
	issn = {1097-0258},
	doi = {10.1002/sim.5907},
	language = {eng},
	number = {30},
	journal = {Statistics in Medicine},
	author = {Haneuse, Sebastien and Rotnitzky, Andrea},
	month = dec,
	year = {2013},
	pmid = {23913589},
	pages = {5260--5277},
}

@article{munoz_population_2012,
	title = {Population {Intervention} {Causal} {Effects} {Based} on {Stochastic} {Interventions}},
	volume = {68},
	copyright = {© 2011, The International Biometric Society},
	issn = {1541-0420},
	url = {https://onlinelibrary.wiley.com/doi/abs/10.1111/j.1541-0420.2011.01685.x},
	doi = {10.1111/j.1541-0420.2011.01685.x},
	language = {en},
	number = {2},
	urldate = {2023-08-29},
	journal = {Biometrics},
	author = {Muñoz, Iván Díaz and van der Laan, Mark},
	year = {2012},
	note = {\_eprint: https://onlinelibrary.wiley.com/doi/pdf/10.1111/j.1541-0420.2011.01685.x},
	pages = {541--549},
}

@article{shimonovich_assessing_2022,
	title = {Assessing the causal relationship between income inequality and mortality and self-rated health: protocol for systematic review and meta-analysis},
	volume = {11},
	issn = {2046-4053},
	shorttitle = {Assessing the causal relationship between income inequality and mortality and self-rated health},
	url = {https://doi.org/10.1186/s13643-022-01892-w},
	doi = {10.1186/s13643-022-01892-w},
	number = {1},
	urldate = {2023-08-29},
	journal = {Systematic Reviews},
	author = {Shimonovich, Michal and Pearce, Anna and Thomson, Hilary and McCartney, Gerry and Katikireddi, Srinivasa Vittal},
	month = feb,
	year = {2022},
	pages = {20},
}

@article{zhou_alcohol_2022,
	title = {Alcohol {Use} and {Use} {Disorder} and {Cancer} {Risk}: {Perspective} on {Causal} {Inference}},
	volume = {8},
	issn = {2673-3005},
	shorttitle = {Alcohol {Use} and {Use} {Disorder} and {Cancer} {Risk}},
	url = {https://doi.org/10.1159/000526407},
	doi = {10.1159/000526407},
	number = {1-2},
	urldate = {2023-08-29},
	journal = {Complex Psychiatry},
	author = {Zhou, Hang and Vasiliou, Vasilis},
	month = aug,
	year = {2022},
	pages = {9--12},
}

@article{stitelman_impact_nodate,
	title = {The {Impact} {Of} {Coarsening} {The} {Explanatory} {Variable} {Of} {Interest} {In} {Making} {Causal} {Inferences}: {Implicit} {Assumptions} {Behind} {Dichotomizing} {Variables}},
	language = {en},
	author = {Stitelman, Ori M and Hubbard, Alan E and Jewell, Nicholas P},
    year = {2010},
	journal = {U.C. Berkeley Division of Biostatistics Working Paper Series},
}

@incollection{rubin_neymans_2015,
	author = {Imbens, Guido W and Rubin, Donald B},
	address = {Cambridge},
	title = {Neyman's {Repeated} {Sampling} {Approach} to {Completely} {Randomized} {Experiments}},
	isbn = {978-0-521-88588-1},
	url = {https://www.cambridge.org/core/books/causal-inference-for-statistics-social-and-biomedical-sciences/neymans-repeated-sampling-approach-to-completely-randomized-experiments/55448ADDB6E99F33966FA4D26F34A79D},
	urldate = {2024-01-27},
	booktitle = {Causal {Inference} for {Statistics}, {Social}, and {Biomedical} {Sciences}: {An} {Introduction}},
	publisher = {Cambridge University Press},
	editor = {Rubin, Donald B. and Imbens, Guido W.},
	year = {2015},
	doi = {10.1017/CBO9781139025751.007},
	pages = {83--112},
}

@article{hu_long-_2015,
	title = {Long- and {Short}-{Term} {Health} {Effects} of {Pesticide} {Exposure}: {A} {Cohort} {Study} from {China}},
	volume = {10},
	issn = {1932-6203},
	shorttitle = {Long- and {Short}-{Term} {Health} {Effects} of {Pesticide} {Exposure}},
	url = {https://www.ncbi.nlm.nih.gov/pmc/articles/PMC4456378/},
	doi = {10.1371/journal.pone.0128766},
	number = {6},
	urldate = {2024-01-27},
	journal = {PLoS ONE},
	author = {Hu, Ruifa and Huang, Xusheng and Huang, Jikun and Li, Yifan and Zhang, Chao and Yin, Yanhong and Chen, Zhaohui and Jin, Yanhong and Cai, Jinyang and Cui, Fang},
	month = jun,
	year = {2015},
	pmid = {26042669},
	pmcid = {PMC4456378},
	pages = {e0128766},
}

@book{rothman_modern_2008,
	address = {Philadelphia, United States},
	title = {Modern {Epidemiology}},
	isbn = {978-1-4698-7382-4},
	url = {http://ebookcentral.proquest.com/lib/berkeley-ebooks/detail.action?docID=3418373},
	urldate = {2024-01-29},
	publisher = {Wolters Kluwer},
	author = {Rothman, Kenneth J. and Greenland, Sander and Lash, Timothy L.},
	year = {2008},
}

@misc{mccoy2023crossvalidated,
      title={Cross-Validated Decision Trees with Targeted Maximum Likelihood Estimation for Nonparametric Causal Mixtures Analysis}, 
      author={David McCoy and Alan Hubbard and Alejandro Schuler and Mark van der Laan},
      year={2023},
      eprint={2302.07976},
      archivePrefix={arXiv},
      primaryClass={stat.ME}
}

@article{robins_estimation_1994,
	title = {Estimation of {Regression} {Coefficients} {When} {Some} {Regressors} are not {Always} {Observed}},
	volume = {89},
	issn = {0162-1459, 1537-274X},
	url = {https://www.tandfonline.com/doi/full/10.1080/01621459.1994.10476818},
	doi = {10.1080/01621459.1994.10476818},
	language = {en},
	number = {427},
	urldate = {2024-02-13},
	journal = {Journal of the American Statistical Association},
	author = {Robins, James M. and Rotnitzky, Andrea and Zhao, Lue Ping},
	month = sep,
	year = {1994},
	pages = {846--866},
}

@article{rubin_estimating_1972,
	title = {Estimating {Causal} {Effects} of {Treatments} in {Experimental} and {Observational} {Studies}},
	volume = {1972},
	copyright = {© 1972 Educational Testing Service},
	issn = {2333-8504},
	url = {https://onlinelibrary.wiley.com/doi/abs/10.1002/j.2333-8504.1972.tb00631.x},
	doi = {10.1002/j.2333-8504.1972.tb00631.x},
	language = {en},
	number = {2},
	urldate = {2024-02-25},
	journal = {ETS Research Bulletin Series},
	author = {Rubin, Donald},
	year = {1972},
	note = {\_eprint: https://onlinelibrary.wiley.com/doi/pdf/10.1002/j.2333-8504.1972.tb00631.x},
	pages = {i--31},
}

@article{stefanski_calculus_2002,
	title = {The {Calculus} of {M}-{Estimation}},
	volume = {56},
	issn = {00031305},
	url = {http://www.jstor.org/stable/3087324},
	number = {1},
	urldate = {2023-11-16},
	journal = {The American Statistician},
	author = {Stefanski, Leonard A. and Boos, Dennis D.},
	year = {2002},
	note = {Publisher: [American Statistical Association, Taylor \& Francis, Ltd.]},
	pages = {29--38},
}

@book{van_der_laan_unified_2003,
	series = {Springer {Series} in {Statistics}},
	title = {Unified {Methods} for {Censored} {Longitudinal} {Data} and {Causality}},
	isbn = {978-0-387-95556-8},
	url = {https://books.google.com/books?id=z4_-dXslTyYC},
	publisher = {Springer},
	author = {van der Laan, Mark J. and Robins, Jamies M.},
	year = {2003},
	lccn = {2002030239},
}

@Article{tmle_r_package,
    title = {{tmle}: An {R} Package for Targeted Maximum Likelihood
      Estimation},
    author = {Susan Gruber and Mark J. {van der Laan}},
    journal = {Journal of Statistical Software},
    year = {2012},
    volume = {51},
    number = {13},
    pages = {1--35},
    url = {https://www.jstatsoft.org/v51/i13/},
    note = {doi:10.18637/jss.v051.i13},
  }

@article{lunceford_stratification_2004,
	title = {Stratification and weighting via the propensity score in estimation of causal treatment effects: a comparative study},
	volume = {23},
	copyright = {Copyright © 2004 John Wiley \& Sons, Ltd.},
	issn = {1097-0258},
	shorttitle = {Stratification and weighting via the propensity score in estimation of causal treatment effects},
	url = {https://onlinelibrary.wiley.com/doi/abs/10.1002/sim.1903},
	doi = {10.1002/sim.1903},
	language = {en},
	number = {19},
	urldate = {2024-05-08},
	journal = {Statistics in Medicine},
	author = {Lunceford, Jared K. and Davidian, Marie},
	year = {2004},
	note = {\_eprint: https://onlinelibrary.wiley.com/doi/pdf/10.1002/sim.1903},
	pages = {2937--2960},
}

@article{schuler_targeted_2017,
	title = {Targeted {Maximum} {Likelihood} {Estimation} for {Causal} {Inference} in {Observational} {Studies}},
	volume = {185},
	issn = {0002-9262},
	url = {https://doi.org/10.1093/aje/kww165},
	doi = {10.1093/aje/kww165},
	number = {1},
	urldate = {2024-05-08},
	journal = {American Journal of Epidemiology},
	author = {Schuler, Megan S. and Rose, Sherri},
	month = jan,
	year = {2017},
	pages = {65--73},
}

@misc{ding_first_2023,
	title = {A {First} {Course} in {Causal} {Inference}},
	url = {http://arxiv.org/abs/2305.18793},
	doi = {10.48550/arXiv.2305.18793},
	urldate = {2024-05-11},
	publisher = {arXiv},
	author = {Ding, Peng},
	month = oct,
    chapter = 10,
	year = {2023},
	note = {arXiv:2305.18793 [stat]},
}

@article{van_der_laan_nonparametric_2023,
	title = {Nonparametric estimation of the causal effect of a stochastic threshold-based intervention},
	volume = {79},
	issn = {1541-0420},
	doi = {10.1111/biom.13690},
	language = {eng},
	number = {2},
	journal = {Biometrics},
	author = {van der Laan, Lars and Zhang, Wenbo and Gilbert, Peter B.},
	month = jun,
	year = {2023},
	pmid = {35526218},
	pmcid = {PMC10024462},
	pages = {1014--1028},
}

@article{diaz_munoz_population_2012,
	title = {Population {Intervention} {Causal} {Effects} {Based} on {Stochastic} {Interventions}},
	volume = {68},
	copyright = {© 2011, The International Biometric Society},
	issn = {1541-0420},
	url = {https://onlinelibrary.wiley.com/doi/abs/10.1111/j.1541-0420.2011.01685.x},
	doi = {10.1111/j.1541-0420.2011.01685.x},
	language = {en},
	number = {2},
	urldate = {2023-07-21},
	journal = {Biometrics},
	author = {Díaz Muñoz, Iván and van der Laan, Mark},
	year = {2012},
	note = {\_eprint: https://onlinelibrary.wiley.com/doi/pdf/10.1111/j.1541-0420.2011.01685.x},
	pages = {541--549},
}

@article{ray_independence_1973,
	title = {Independence of {Irrelevant} {Alternatives}},
	volume = {41},
	issn = {0012-9682},
	url = {https://www.jstor.org/stable/1913820},
	doi = {10.2307/1913820},
	number = {5},
	urldate = {2024-05-21},
	journal = {Econometrica},
	author = {Ray, Paramesh},
	year = {1973},
	note = {Publisher: [Wiley, Econometric Society]},
	pages = {987--991},
}

@article{phillips_practical_2023,
	title = {Practical considerations for specifying a super learner},
	volume = {52},
	issn = {0300-5771},
	url = {https://doi.org/10.1093/ije/dyad023},
	doi = {10.1093/ije/dyad023},
	number = {4},
	urldate = {2024-05-23},
	journal = {International Journal of Epidemiology},
	author = {Phillips, Rachael V and van der Laan, Mark J and Lee, Hana and Gruber, Susan},
	month = aug,
	year = {2023},
	pages = {1276--1285},
}

@article{chernozhukov_doubledebiased_2018,
	title = {Double/debiased machine learning for treatment and structural parameters},
	volume = {21},
	issn = {1368-4221},
	url = {https://doi.org/10.1111/ectj.12097},
	doi = {10.1111/ectj.12097},
	number = {1},
	urldate = {2024-05-23},
	journal = {The Econometrics Journal},
	author = {Chernozhukov, Victor and Chetverikov, Denis and Demirer, Mert and Duflo, Esther and Hansen, Christian and Newey, Whitney and Robins, James},
	month = feb,
	year = {2018},
	pages = {C1--C68},
}

@article{rubin_randomization_1980,
	title = {Randomization {Analysis} of {Experimental} {Data}: {The} {Fisher} {Randomization} {Test} {Comment}},
	volume = {75},
	issn = {0162-1459},
	shorttitle = {Randomization {Analysis} of {Experimental} {Data}},
	url = {https://www.jstor.org/stable/2287653},
	doi = {10.2307/2287653},
	number = {371},
	urldate = {2024-07-08},
	journal = {Journal of the American Statistical Association},
	author = {Rubin, Donald B.},
	year = {1980},
	note = {Publisher: [American Statistical Association, Taylor \& Francis, Ltd.]},
	pages = {591--593},
}

@article{nugent_demonstration_2023,
	title = {A {Demonstration} of {Modified} {Treatment} {Policies} to {Evaluate} {Shifts} in {Mobility} and {COVID}-19 {Case} {Rates} in {US} {Counties}},
	volume = {192},
	issn = {1476-6256},
	doi = {10.1093/aje/kwad005},
	language = {eng},
	number = {5},
	journal = {American Journal of Epidemiology},
	author = {Nugent, Joshua R. and Balzer, Laura B.},
	month = may,
	year = {2023},
	pmid = {36623841},
	pages = {762--771},
}

@article{van_der_laan_unified_2003_cv,
	title = {Unified {Cross}-{Validation} {Methodology} {For} {Selection} {Among} {Estimators} and a {General} {Cross}-{Validated} {Adaptive} {Epsilon}-{Net} {Estimator}: {Finite} {Sample} {Oracle} {Inequalities} and {Examples}},
	shorttitle = {Unified {Cross}-{Validation} {Methodology} {For} {Selection} {Among} {Estimators} and a {General} {Cross}-{Validated} {Adaptive} {Epsilon}-{Net} {Estimator}},
	url = {https://biostats.bepress.com/ucbbiostat/paper130},
	journal = {U.C. Berkeley Division of Biostatistics Working Paper Series},
	author = {van der Laan, Mark and Dudoit, Sandrine},
	month = nov,
	year = {2003},
}

@article{diaz_nonparametric_2023,
	title = {Nonparametric {Causal} {Effects} {Based} on {Longitudinal} {Modified} {Treatment} {Policies}},
	volume = {118},
	issn = {0162-1459},
	url = {https://doi.org/10.1080/01621459.2021.1955691},
	doi = {10.1080/01621459.2021.1955691},
	number = {542},
	urldate = {2024-08-22},
	journal = {Journal of the American Statistical Association},
	author = {Díaz, Iván and Williams, Nicholas and Hoffman, Katherine L. and Schenck, Edward J.},
	month = apr,
	year = {2023},
	note = {Publisher: ASA Website
\_eprint: https://doi.org/10.1080/01621459.2021.1955691},
	pages = {846--857},
}

@incollection{hoffman_chapter_2019,
	title = {Chapter 20 - {Odds} {Ratio}, {Relative} {Risk}, {Attributable} {Risk}, and {Number} {Needed} to {Treat}},
	isbn = {978-0-12-817084-7},
	url = {https://www.sciencedirect.com/science/article/pii/B9780128170847000206},
	urldate = {2025-01-31},
	booktitle = {Basic {Biostatistics} for {Medical} and {Biomedical} {Practitioners} ({Second} {Edition})},
	publisher = {Academic Press},
	author = {Hoffman, Julien I. E.},
	editor = {Hoffman, Julien I. E.},
	month = jan,
	year = {2019},
	doi = {10.1016/B978-0-12-817084-7.00020-6},
	pages = {295--310},
}

@article{tran_et_al,
author = {Kathy V. Tran  and Joan A. Casey  and Lara J. Cushing  and Rachel Morello-Frosch },
title = {Residential Proximity to Oil and Gas Development and Birth Outcomes in California: A Retrospective Cohort Study of 2006–2015 Births},
journal = {Environmental Health Perspectives},
volume = {128},
number = {6},
pages = {067001},
year = {2020},
doi = {10.1289/EHP5842},

URL = {https://ehp.niehs.nih.gov/doi/abs/10.1289/EHP5842},
eprint = {https://ehp.niehs.nih.gov/doi/pdf/10.1289/EHP5842}

}

@electronic{
 ca_sb_1137,
 title = {SB1137 \emph{Oil and gas: operations: location restrictions: notice of intention: health protection zone: sensitive receptors.}},
author = {{California Senate, 2022 Session}},
 number = {SB 1137},
 session = {2021-2022},
chapter = {365},
code={Public Resources},
month = {September 16},
year = {2022},
url = {https://leginfo.legislature.ca.gov/faces/billNavClient.xhtml?bill_id=202120220SB1137}
}
\bibliographystyle{vancouver}

\newpage
\appendix
\appendixpage
\addappheadtotoc
\section{Estimators}
\label{appendix_est}

\subsection{Regression Estimator}
\label{appendix_reg}
The regression estimator $\widehat{\Psi}^{reg}$ first requires researchers to calculate the following ordinary least squares (OLS) coefficient estimates with $Y$ as the outcome:

$$\hat{Y} = \hat{\beta_0} +  T \hat{\beta}_{T}+ \widecheck{W}'\hat{\beta}_W + T \widecheck{W}'\hat{\beta}_{interact},$$
where $\widecheck{W} = W - \frac{1}{n}\sum_{i=1}^n W_i$ is the vector of demeaned confounders. The regression estimator will be consistent when the regression estimate $\hat{Y}$ given the binarized exposure $T$ and the demeaned confounders $\widecheck{W}$ is consistent for $Y$. For a more thorough treatment of this topic, we refer the reader to Appendix \ref{appendix_consistency}. 

Let $\overline{T} = \frac{1}{n}\sum_{i=1}^n T_i$ and $\overline{T \widecheck{W}'} = \frac{1}{n}\sum_{i=1}^n T_i  \widecheck{W}'$ be the mean of $T$ and $T \widecheck{W}'$ in the sample, respectively. Then, the regression estimates for $\Psi_{BATE}$ and $\Psi_{CAB}$ are given by 

\begin{align*}
    \widehat{\Psi}_{BATE}^{reg} &= \hat{\beta}_T\\ \widehat{\Psi}_{CAB}^{reg} &= \begin{cases}
        (1-\overline{T})\hat{\beta}_T - \overline{T \widecheck{W}'}\hat{\beta}_{interact}  \text{ for } t = 1\\
        (-\overline{T})\hat{\beta}_T - \overline{T \widecheck{W}'}\hat{\beta}_{interact}  \text{ for } t = 0\\
    \end{cases},\\
\end{align*}
respectively.

For $\widehat{\Psi}_{BATE}^{reg}$ and $\widehat{\Psi}_{CAB}^{reg}$, the respective variance estimators are given by

\begin{align*}
    \widehat{\mathbb{V}}\left(\widehat{\Psi}_{BATE}^{reg}\right) &= 
    \begin{bmatrix}
    0 & 0 & 0 & 1 & 0 & 0\\
    \end{bmatrix} \mathbb{V}(\hat{\beta}) 
    \begin{bmatrix}
    0 \\ 0 \\ 0 \\ 1 \\ 0 \\ 0\\
    \end{bmatrix}\\
    \widehat{\mathbb{V}}\left(\widehat{\Psi}_{CAB}^{reg}\right) &= \begin{cases}
            \begin{bmatrix}
    - \hat{\beta_T} & -\hat{\beta}_{interact} & 0 & (1-\overline{T}) & 0 & -\overline{T \widecheck{W}'}
    \end{bmatrix} \widehat{\mathbb{V}}(\hat{\beta}) \begin{bmatrix}
    - \hat{\beta_T} \\ -\hat{\beta}_{interact} \\ 0 \\ (1-\overline{T}) \\ 0 \\ -\overline{T \widecheck{W}}
    \end{bmatrix} \text{ for } t = 1\\
        \begin{bmatrix}
    - \hat{\beta_T} & -\hat{\beta}_{interact} & 0 & -\overline{T} & 0 & -\overline{T \widecheck{W}'}
    \end{bmatrix} \widehat{\mathbb{V}}(\hat{\beta}) \begin{bmatrix}
    - \hat{\beta_T} \\ -\hat{\beta}_{interact} \\ 0 \\ -\overline{T} \\ 0 \\ -\overline{T \widecheck{W}}
    \end{bmatrix} \text{ for } t = 0
    \end{cases},
\end{align*}

\noindent where $r_i = Y_i - (\hat{\beta_0} + T_i \hat{\beta}_{T} + 
\widecheck{W_i}'\hat{\beta}_W + T_i \widecheck{W}_i'\hat{\beta}_{interact})$ is the residual from the regression and $\mathbb{V}(\widehat{\theta})$ is the variance estimator of the vector of M-estimators $\widehat{\theta} = (\overline{T}, \overline{T \widecheck{W}}, \hat{\beta_0}, \hat{\beta}_{T}, \hat{\beta}_{W}, \hat{\beta}_{interact})'$ given by 

\begin{align*}
    \widehat{\mathbb{V}}(\widehat{\theta}) &= A_n(\widehat{\theta})^{-1} B_n(\widehat{\theta}) (A_n(\widehat{\theta})^{-1})',
\end{align*}
where 
\begin{align*}
    \psi(O_i, \hat{\theta}) &= \begin{bmatrix}
        T_i - \overline{T}\\
        T_i\widecheck{W}_i - \overline{T \widecheck{W}}\\
        r_i\\
        r_iT_i\\
        r_i\widecheck{W}_i\\
        r_iT_i\widecheck{W}_i
    \end{bmatrix}\\
    C_i &= \begin{bmatrix}
        1 & T_i & \widecheck{W_i}' & T_i\widecheck{W_i}'        
    \end{bmatrix} \begin{bmatrix}
        1\\
        T_i\\
        \widecheck{W_i}\\
        T_i\widecheck{W_i}\\
    \end{bmatrix}\\
    A_n(\widehat{\theta}) &= \frac{1}{n} \sum_{i=1}^n \begin{bmatrix}
        1 & 0 & 0  \\
        0 & 1 & 0 \\
        0 & 0 & C_i\\
    \end{bmatrix}\\
    B_n(\widehat{\theta}) &= \frac{1}{n} \sum_{i=1}^n \psi(O_i, \widehat{\theta})\psi(O_i, \widehat{\theta})'.\\
\end{align*}

\subsection{Inverse Probability Weighting (IPW) Estimator}
\label{appendix_ipw}

The IPW estimator $\widehat{\Psi}^{IPW}$ requires an estimator of the propensity score $\pi_\mathcal{A}(w) = Pr(T=1|W=w)$ for all $w \in \text{supp}(W)$. If this estimator is consistent for the true propensity score, then the IPW estimator will be consistent. Denote this consistent estimate $\widehat{\pi_\mathcal{A}}(w)$. Then, the IPW estimators for $\Psi_{BATE}$ and $\Psi_{CAB}$ are given by

\begin{align}
    \widehat{\Psi}_{BATE}^{IPW} &= \frac{1}{n} \left(\sum_{i=1}^n \frac{T_iY_i}{\widehat{\pi_\mathcal{A}}(w)} - \sum_{i=1}^n \frac{(1-T_i)Y_i}{1-\widehat{\pi_\mathcal{A}}(w)}\right)\\
    \widehat{\Psi}_{CAB}^{IPW} &= \renewcommand{\arraystretch}{1.5}\left\{\begin{array}{@{}l@{\quad}l@{}}
        \cfrac{1}{n} \left(\mathlarger{\sum}_{i=1}^n \cfrac{T_iY_i}{\widehat{\pi_\mathcal{A}}(w)} -  \mathlarger{\sum}_{i=1}^n Y_i\right) \text{ for } t=1\\
        \cfrac{1}{n} \left(\mathlarger{\sum}_{i=1}^n \cfrac{(1-T_i)Y_i}{1-\widehat{\pi_\mathcal{A}}(w)} -  \mathlarger{\sum}_{i=1}^n Y_i\right) \text{ for } t=0\\     
    \end{array}\right.\kern-\nulldelimiterspace,
\end{align}
respectively. The IPW estimators is asymptotically linear and therefore converges in distribution to a normal distribution \cite{van_der_laan_unified_2003}. 

One could estimate the standard deviation of this IPW estimator using the influence function of the estimator \textit{assuming the propensity score $\pi_{\mathcal{A}}(w)$ were known}. However, such estimates are too large and thus result in overly-conservative inference \cite{lunceford_stratification_2004}. Therefore, we suggest that researchers calculate the standard errors for the IPW estimators described here via the nonparametric bootstrap.

\label{appendix_AIPW}
\subsection{Augmented IPW Estimator (AIPW)}

The AIPW estimator $\widehat{\Psi}^{AIPW}$ is a semi-parametrically efficient, doubly-robust estimator. $\widehat{\Psi}^{AIPW}$ requires both an estimator of the conditional mean $m(t,w) \equiv E[Y|t, w]$ and an estimator of the propensity score conditional on confounders $\pi_\mathcal{A}(w)$. Denote these estimates $\widehat{m}(t,w)$ and $\widehat{\pi_\mathcal{A}}(w)$, respectively. Researchers can use a variety of machine learning (ML) algorithms to create these estimates, providing more flexibility. Super learning can also be used to determine the optimal predictor (or ensemble of predictors) from a class of many candidate estimators based on the data \cite{phillips_practical_2023}. Researchers should employ sample splitting and cross fitting to reduce over-fitting and ensure that the estimates converge to the truth sufficiently quickly \cite{van_der_laan_unified_2003_cv, chernozhukov_doubledebiased_2018}.

The AIPW estimators for $\Psi_{BATE}$ and $\Psi_{CAB}$ are given by

\begin{align}
    \widehat{\Psi}_{BATE}^{AIPW} &= \frac{1}{n} \left[ \sum_{i=1}^n \left(\frac{T_i}{\widehat{\pi_\mathcal{A}}(w)} - \frac{(1-T_i)}{1-\widehat{\pi_\mathcal{A}}(w)}\right)(Y_i - \widehat{m}(T_i, W_i)) + \widehat{m}(1,W_i) - \widehat{m}(0,W_i)\right]\\
    \widehat{\Psi}_{CAB}^{AIPW} &= \renewcommand{\arraystretch}{1.5}\left\{\begin{array}{@{}l@{\quad}l@{}}
        \cfrac{1}{n} \left[\mathlarger{\sum}_{i=1}^n \cfrac{T_i}{\widehat{\pi_\mathcal{A}}(w)}(Y_i-\widehat{m}(1, W_i)) + \widehat{m}(1,W) - Y_i\right] \text{ for } t=1\\
        \cfrac{1}{n} \left[\mathlarger{\sum}_{i=1}^n \cfrac{(1-T_i)}{1-\widehat{\pi_\mathcal{A}}(w)}(Y_i-\widehat{m}(0,W_i)) + \widehat{m}(0,W_i) - Y_i\right] \text{ for } t=0\\     
    \end{array}\right.\kern-\nulldelimiterspace,
\end{align}

\noindent respectively.

The variance of the two estimators can be calculated either via estimation of the influence curve or via nonparametric bootstrap. 

The influence curve evaluated at a given observation $O_i$ for $\widehat{\Psi}_{BATE}^{AIPW}$ and $\widehat{\Psi}_{CAB}^{AIPW}$ can be estimated as

\begin{align*}
    \widehat{\phi}(O_i)_{BATE} &=  \left(\frac{T_i}{\widehat{\pi_\mathcal{A}}(w)} - \frac{(1-T_i)}{1-\widehat{\pi_\mathcal{A}}(w)}\right)(Y - \widehat{m}(T_i, W_i)) + \widehat{m}(1,W_i) - \widehat{m}(0,W_i) - \widehat{\Psi}_{BATE}\\
    \widehat{\phi}(O_i)_{CAB} &= \renewcommand{\arraystretch}{1.5}\left\{\begin{array}{@{}l@{\quad}l@{}}
        \cfrac{T_i}{\widehat{\pi_\mathcal{A}}(w)}(Y_i-\widehat{m}(1, W_i)) + \widehat{m}(1,W) - Y_i- \widehat{\Psi}_{CAB}^{AIPW}  \text{ for } t=1\\
        \cfrac{(1-T_i)}{1-\widehat{\pi_\mathcal{A}}(w)}(Y_i-\widehat{m}(0,W_i)) + \widehat{m}(0,W_i) - Y_i - \widehat{\Psi}_{CAB}^{AIPW} \text{ for } t=0.\\     
    \end{array}\right.\kern-\nulldelimiterspace
\end{align*}

\noindent Then, the variance estimates for $\widehat{\Psi}_{BATE}^{AIPW}$ and $\widehat{\Psi}_{CAB}^{AIPW}$ can be calculated as the estimated variance of the influence curve divided by the sample size $n$:

\begin{align*}
\widehat{\mathbb{V}}\left(\widehat{\Psi}_{BATE}^{AIPW}\right) &= \frac{1}{n} \widehat{\text{Var}} \left( \widehat{\phi}(O_i)_{BATE}\right)\\
\widehat{\mathbb{V}}\left(\widehat{\Psi}_{CAB}^{AIPW}\right) &= \frac{1}{n} \widehat{\text{Var}} \left( \widehat{\phi}(O_i)_{CAB} \right).\\
\end{align*}

\subsection{Targeted Maximum Likelihood Estimator (TMLE)}
\label{appendix_tmle}

The TMLE estimator $\widehat{\Psi}^{TMLE}$ is another semi-parametrically efficient, doubly-robust estimator. For a more comprehensive introduction to TMLE for applied work in the binary exposure setting, we refer researchers to the 2017 paper ``Targeted Maximum Likelihood Estimation for Causal Inference in Observational Studies'' by Schuler and Rose \cite{schuler_targeted_2017}. The TMLE presented for the BATE is equivalent to the binTMLE presented in \cite{van_der_laan_nonparametric_2023} and achieves the semi-parametric efficiency bound under the assumption of missingness at random. We provide a brief overview of the steps required below.

\begin{enumerate}

    \item Estimate the conditional expectation function $E[Y|T=t, W=w]$ for $t \in \{0,1\}, w \in \text{supp}(W)$. Use this estimator to generate initial estimates $\widehat{Y}_{\tilde A_1}$ and $\widehat{Y}_{\tilde{A}_0}$ for each individual in the sample, which are the predicted expectations when $T=1$ and $T=0$, respectively. Let $\hat{Y} = 1(T=1) \hat{Y}_{\tilde A_1} + 1(T=0) \hat{Y}_{\tilde A_0}$ be the prediction for the observed outcome.
    \item Estimate the propensity score function $Pr(T=1|W=w)$ for $w \in \text{supp}(W)$. Denote this estimator $g_0(w)$. 
    \item Calculate the ``clever covariate'' $H(T=t, W=w)$, which differs for the two estimators,  for each individual in the sample.
    \begin{itemize}
        \item If estimating $\Psi_{BATE}$, $H = \frac{1(T=1)}{g_0(W)} - \frac{1(T=0)}{1-g_0(W)}$.
        \item If estimating $\Psi_{CAB}$, $H = \frac{1(T=1)}{g_0(W)}-1$ when considering $E[Y_{\tilde{A}_1}]$ and $H = \frac{1(T=0)}{1-g_0(W)}-1$ when considering $E[Y_{\tilde{A}_0}]$.
    \end{itemize}
    Then, use logistic regression to estimate the coefficient $\beta$ regressing H on the observered outcome $Y$ using $\hat{Y}$ as a fixed intercept:
    $$\text{logit}(Y) = \text{logit}(\hat{Y}) + \beta H(T,W).$$
    Use this estimated coefficient $\hat{\beta}$ to calculate new, ``updated" predictions for the potential outcomes $\hat{Y}_{\tilde{A}_1}^*$ and $\hat{Y}_{\tilde{A}_0}^*$ for each individual in the sample:
    \begin{align*}
        \text{logit}(\hat{Y}_{\tilde{A}_1}^*) &= \text{logit}(\widehat{Y}_{\tilde A_1}) + \hat\beta \frac{1}{g_0(W)}\\
        \text{logit}(\hat{Y}_{\tilde{A}_0}^*) &= \text{logit}(\widehat{Y}_{\tilde A_0}) + \hat\beta \frac{1}{1-g_0(W)}.
    \end{align*}
    \item Let $\hat{Y}^* = 1(T=1) \hat{Y}_{\tilde{A}_1}^* + 1(T=0) \hat{Y}_{\tilde{A}_0}^*$ be the new prediction for the observed outcome. Now, we can calculate
    \begin{align*}
        \hat{\Psi}_{BATE}^{TMLE} &= \frac{1}{n} \sum_{i=1}^n \hat{Y}_{\tilde{A}_1}^* - \hat{Y}_{\tilde{A}_0}^* \\
        \hat{\Psi}_{CAB}^{TMLE} &= \frac{1}{n} \sum_{i=1}^n \hat{Y}_{\tilde{A}_t}^* - \hat{Y}^*.
    \end{align*}
\end{enumerate}

To estimate the variance, researchers can either calculate influence curve based estimates or use the nonparametric bootstrap. The influence curve based estimates are the same as the AIPW estimator as in Appendix \ref{appendix_AIPW} and have the same consistency conditions.

If estimating $\Psi_{BATE}$, researchers can use the R package \texttt{tmle}, using $T$ as the binary exposure variable \cite{tmle_r_package}. The package will calculate the estimate as well as the standard errors based on the influence curve.

\section{When is the regression estimator consistent?}
\label{appendix_consistency}

Suppose the expectation of $Y$ given $A$ and $W$ is linear in $W$ and $W$ times some function of $A$. In order for the regression estimator to be consistent, the expectation of $Y$ given $T$ and $W$ must be linear. When will this be the case? Let $g_A(A)$ and $g_{int}(A)$ be two (potentially non-linear) functions of $A$. 

Consider a general DGP where the expectation of $Y$ given $A$ and $W$ is a linear function in $W$, $g_A(A)$, and in an interaction term $W*g_{int}(A)$, i.e.

$$E[Y|A,W] = \beta_0 + \beta_W W + \beta_A g_A(A) + \beta_{int} W * g_{int}(A).$$

Consider a cut-off at $A=a_0$ such that $T=1$ when $A \geq a_0$ and $T=0$ when $A < a_0$. Then, we can consider the expectation of $Y$ given $T=1$ and $W$:
\begin{align*}
    E[Y|T=1, W] &= \frac{1}{\pi_{\mathcal{A}}(w)} \int_{a_0}^\infty E[Y|A,W] da \\
    &= \frac{1}{\pi_{\mathcal{A}}(w)} \int_{a_0}^\infty \beta_0 + \beta_W W + \beta_A g_A(a) + \beta_{int} W * g_{int}(a)da\\
    &=\frac{1}{\pi_{\mathcal{A}}(w)} \left( \beta_0 + \beta_W W\right) + \frac{\beta_A}{\pi_{\mathcal{A}}(w)}\int_{a_0}^\infty g_A(a)da +\beta_{int}* W \frac{1}{\pi_{\mathcal{A}}(w)}\int_{a_0}^\infty g_{int}(a)da\\
    &= \frac{1}{\pi_{\mathcal{A}}(w)} \left( \beta_0 + \beta_W W\right) + \beta_A E[g_A(\tilde{A}_1)|W] + \beta_{int} W  E[g_{int}(\tilde{A}_1)|W].\\
\end{align*}

\noindent Similarly, for $E[Y|T=0, W]$, we have

$$E[Y|T=0, W] = \frac{1}{1-\pi_{\mathcal{A}}(w)} \left( \beta_0 + \beta_W W\right) + \beta_A E[g_A(\tilde{A}_0)|W] + \beta_{int} W  E[g_{int}(\tilde{A}_0)|W].$$

Therefore, the regression estimator for $\Psi_{BATE}$ will be consistent when $E[g_A(\tilde{A}_t)|W]$ and $E[g_{int}(\tilde{A}_t)|W]$ are linear in $W$ for $t \in \{0,1\}$. The regression estimator for $\Psi_{CAB}$ when considering $T=t$ will be consistent when $E[g_A(\tilde{A}_t)|W]$ and $E[g_{int}(\tilde{A}_t)|W]$ are linear in $W$.

\end{document}